\documentclass[twocolumn,aps,showpacs,prb,tightenlines,amsmath,amssymb]{revtex4}
\usepackage{graphicx}
\usepackage{amssymb}
\usepackage{dcolumn}
\usepackage{amsmath}
\usepackage{pifont}
\usepackage{bm}
\usepackage{colordvi}
\newcommand{\bgreek}[1]{\mbox{\boldmath$#1$\unboldmath}}

\begin{document}

\title{Electron spin relaxation due to D'yakonov-Perel' and
  Elliot-Yafet mechanisms in monolayer MoS$_2$:\\Role of intravalley
  and intervalley processes}
\author{L. Wang}
\affiliation{Hefei National Laboratory for Physical Sciences at
Microscale and Department of Physics,
University of Science and Technology of China, Hefei,
Anhui, 230026, China}
\author{M. W. Wu}
\thanks{Author to whom correspondence should be addressed}
\email{mwwu@ustc.edu.cn.}
\affiliation{Hefei National Laboratory for Physical Sciences at
Microscale and Department of Physics, University of Science and
Technology of China, Hefei, Anhui, 230026, China}

\date{\today}

\begin{abstract}
We investigate the in-plane spin relaxation of electrons due to the D'yakonov-Perel' and
Elliot-Yafet mechanisms including the intra- and inter-valley processes
in monolayer MoS$_2$. We construct the effective Hamiltonian for the
conduction band using the L\"{o}wdin partition method from 
the anisotropic two-band Hamiltonian with the intrinsic
spin-orbit coupling of the conduction band included. The spin-orbit
coupling of the conduction band induces the intra- and
inter-valley D'yakonov-Perel' spin relaxation. In addition, the
  Elliot-Yafet spin relaxation also takes place due to the interband
  spin mixing. We find that the
D'yakonov-Perel' mechanism dominates the in-plane spin relaxation. 
In the framework of this mechanism, the intravalley process is shown to play a more important
role at low temperature whereas the intervalley one becomes more important
at high temperature. At the temperature in between, the leading process of the
in-plane spin relaxation changes from the intervalley to intravalley
one as the electron density increases. 
Moreover, we find that the intravalley process
is dominated by the electron-electron Coulomb scattering even with
high impurity density since the dominant term in
the spin-orbit coupling is isotropic, which does not lead to the spin
relaxation together with the electron-impurity scattering. This is very different from the previous studies in
semiconductors and graphene.
\end{abstract}


\maketitle
\section{INTRODUCTION}
In the past few years, great efforts have been devoted to a new two-dimensional
material, i.e., monolayer
MoS$_2$\cite{novoselov102,matte122,radisavljevic6} since it has a
direct gap at K(K$^{\prime}$)
point,\cite{splendiani10,kaasbjerg,xli,zhu84,chei,hshi,xiao108,yun85,rostami,zahid,kormanyos,cappelluti,mak105,
eda11,korn99,ellis99,wang7}
valley-dependent interband optical selection
rule,\cite{cao3,mak7,zeng7,xiao108,sallen86,lagarde} and also good 
spin properties.\cite{zhu84,xiao108,chei,kadantsev152,ochoa,hshi,rostami,zahid,kormanyos,cappelluti,kosmider87}
Specifically, both the conduction and valence bands are spin splitted
caused by the space inversion
asymmetry.\cite{zhu84,ochoa,xiao108,kormanyos,zahid,chei,kadantsev152,kosmider87,hshi,rostami,cappelluti}
This spin splitting at K(K$^{\prime}$) point has opposite signs due to
the time reversal symmetry, which makes the spintronics intriguing in this
multivalley system.

Spin relaxation, which is one of the prerequisites in realizing spintronic devices, has been
studied in monolayer MoS$_2$ very recently.\cite{ochoa,wangarXiv} 
There exist both the intravalley\cite{ochoa} and
intervalley\cite{wangarXiv} spin relaxation processes. 
Specifically, Ochoa and Rold\'{a}n\cite{ochoa} investigated the intravalley
electron relaxation process for in-plane spins 
due to what they claimed \Green{both the} D'yakonov-Perel'\cite{dp} (DP) and
Elliot-Yafet\cite{ey} (EY) mechanisms. However, the intravalley DP
spin relaxation process contributed by the disorder is in fact absent
since the isotropic low-energy two-band
Hamiltonian,\cite{xiao108} i.e., 
\begin{eqnarray}
H^{\rm i}_{\mu}=\Delta\tau_z/2+\lambda_v\mu\sigma_z(1-\tau_z)/{2}+t_0a_0{\bf k}\cdot{\bgreek \tau}_{\mu},\label{xiaohamil}
\end{eqnarray}
is employed in their calculation.\cite{dp,wureview} Here, $\bgreek \tau$ and
$\bgreek \sigma$ are the Pauli matrices for two Bloch basis
functions and spins, respectively; ${\bgreek \tau}_{\mu}=(\mu\tau_x,\tau_y)$; valley index
$\mu=1(-1)$ represents K(K$^{\prime}$) valley; $\Delta$ is the
energy gap; $\lambda_v$ is the strength of the intrinsic spin-orbit coupling (SOC) of the valence band.
This isotropic Hamiltonian can 
lead to the intravalley spin relaxation in the presence of the
electron-electron Coulomb and intravalley electron-phonon
scatterings. These scatterings are
unfortunately  absent in Ref.~\onlinecite{ochoa}.  In
addition, according to the latest report by Rostami {\it et
  al.},\cite{rostami} the isotropic two-band Hamiltonian becomes anisotropic
when higher order terms in the momentum are taken into
account. Specifically, 
\begin{eqnarray}
H^{\rm a}_{\mu}&=&H^{\rm i}_{\mu}+\frac{\hbar^2k^2}{4m_0}(\alpha+\beta\tau_z)+t_1a_0^2{\bf k}\cdot{\bgreek
  \tau}_{\mu}^*\tau_x{\bf k}\cdot{\bgreek \tau}_{\mu}^*,\label{totalhamil}
\end{eqnarray}
in which the last term is anisotropic. This anisotropic term can 
cause the DP spin relaxation together with the
electron-impurity scattering. 
Moreover, the intrinsic SOC of the
conduction band [$\lambda_c\mu \sigma_z(1+\tau_z)/{2}$], which provides opposite effective magnetic fields in the
two valleys,\cite{kormanyos,kormanyos2} is neglected in the above
Hamiltonians [Eqs.~(\ref{xiaohamil}) and (\ref{totalhamil})]. This SOC has been
demonstrated to open an intervalley DP relaxation process for in-plane
spins in the presence of intervalley electron-phonon scattering 
very recently by Wang and Wu.\cite{wangarXiv} Therefore, it is of crucial
importance to give a full investigation on the spin relaxation 
 and compare the relative importance of each mechanism in
monolayer MoS$_2$.

In the present work, by taking into account all the relevant scatterings, 
we study the electron spin relaxation due to the DP and EY mechanisms with the intra- and
inter-valley processes included in monolayer MoS$_2$ by 
the kinetic spin Block equation (KSBE) approach.\cite{wureview} 
With the anisotropic two-band Hamiltonian by Rostami {\it et
    al.}\cite{rostami} [see Eq.~(\ref{totalhamil})] and the intrinsic
  SOC of the conduction band included,\cite{kormanyos,kormanyos2} 
the SOC of the conduction band near the K(K$^{\prime}$) point to the third order of the momentum is given by
\begin{equation}
{\bf \Omega}^{\mu}=[2\lambda_c\mu+\mu A_1k^2+A_2(k^3_x-3k_xk_y^2)]\hat{\bf z},
\label{soc}
\end{equation}
which is obtained by the L\"{o}wdin partition method.\cite{lowdin,winkler}
Here, the $z$-axis is set to be out of the monolayer MoS$_2$ plane; the
coefficients $A_1$ and $A_2$ are given in Appendix~\ref{appA}. The
first two terms contain valley index and therefore provide an intervalley inhomogeneous
broadening\cite{wuning} for in-plane spins, which leads to intervalley DP spin relaxation together
with the intervalley scattering.\cite{pzhang112,lwang,wangarXiv} In addition,
the last two
terms are momentum dependent, which induce not only the intervalley DP spin
relaxation with the intervalley scattering but also the intravalley one with the
intravalley scattering. It is noted that only the anisotropic cubic term
causes the DP spin relaxation with the
electron-impurity scattering. 

In addition to the DP mechanism, we also include the EY one. However, the contribution
of the EY mechanism to the in-plane spin relaxation is negligible compared with that of
the DP one due to the marginal
in-plane spin mixing. In the framework of the DP mechanism, we find that the intravalley
process plays a more important role at low temperature. However, at high
temperature, the intervalley process becomes more important. As for the
temperature in between, we find that the leading process of the spin
relaxation changes from the intervalley to
intravalley one with increasing electron density. Moreover, we
  find that the electron-electron Coulomb scattering dominates the intravalley
process even in the presence of high impurity density, due to the
negligible inhomogeneous broadening from the anisotropic
 term in the SOC. This is very
 different from semiconductors\cite{wureview} and graphene.\cite{yzhou,lwang}

This paper is organized as follows. In Sec.~II, we introduce our model and the
numerical method. Then in Sec.~III, we investigate the temperature and electron-density dependences of
the in-plane  spin relaxation. We summarize in Sec.~IV.

\section{MODEL AND KSBEs}
In monolayer MoS$_2$, the effective Hamiltonian of the conduction band 
near the K(K$^{\prime}$) point can be written as
\begin{eqnarray}
H_{\rm eff}^{\mu}&=&\epsilon_{\mu{\bf k}}+{\bf \Omega}^{\mu}\cdot{\bgreek \sigma}/2\label{hamil}
\end{eqnarray}
where $\epsilon_{\mu{\bf k}}=\hbar^2{\bf k}^2/(2m^*)$ with $m^*$ being the
effective mass and ${\bf \Omega}^{\mu}$ is given in Eq.~(\ref{soc}). This
effective Hamiltonian is obtained by the L\"{o}wdin partition method\cite{lowdin,winkler}
up to the third order of the momentum.\cite{warping} The details are shown in
Appendix~\ref{appA}. It is noted that similar effective Hamiltonian with the
cubic term neglected has been given by Korm\'{a}nyos {\it et al.}.\cite{kormanyos2}

We then construct the microscopic KSBEs\cite{wureview} to study the 
electron spin relaxation in monolayer MoS$_2$. The KSBEs read\cite{wureview} 
\begin{eqnarray}
\partial_t{\rho}_{\mu{\bf k}}=\partial_t{\rho}_{\mu{\bf k}}|_{\rm coh}+\partial_t{\rho}_{\mu{\bf k}}|_{\rm scat},
\label{KSBE}
\end{eqnarray}  
where ${\rho}_{\mu{\bf k}}$ stand for the density matrices of electrons
with the diagonal terms $\rho_{\mu{\bf k},\sigma\sigma}\equiv f_{\mu{\bf k}\sigma}\ (\sigma=\pm \frac{1}{2})$
denoting the distribution functions and the off-diagonal ones $\rho_{\mu{\bf k},(\frac{1}{2})(-\frac{1}{2})}
=\rho_{\mu{\bf k},(-\frac{1}{2})(\frac{1}{2})}^*$ being the spin
coherence. The coherent terms $\partial_t{\rho}_{\mu{\bf k}}|_{\rm
    coh}$ are given in Ref.~\onlinecite{yzhou}. 
$\partial_t{\rho}_{\mu{\bf k}}|_{\rm scat}$ are the scattering terms 
including the spin conserving terms, i.e., the electron-electron Coulomb, 
electron-impurity, intravalley electron-acoustic phonon, electron-optical
phonon, and also the intervalley electron-phonon\cite{kaasbjerg,wangarXiv} (electron-KTA, -KLA,
-KTO, and -KLO) scatterings and the spin-flip
terms due to the EY mechanism.\cite{ey} Their detailed expressions can be found in
Ref.~\onlinecite{jiang}\cite{scat} and here we only show the
  electron-electron Coulomb scattering in Appendix~\ref{appA}. The scattering matrix
elements have been introduced in Ref.~\onlinecite{wangarXiv} and the spin mixing
$\hat{\Lambda}_{\mu{\bf k},{\mu}^{\prime}{\bf k}^{\prime}}$ for the conduction
band in the spin-flip scattering terms is given in Appendix~\ref{appA}.

\section{NUMERICAL RESULTS}
In the calculation, the effective mass, calculated from
Eq.~(\ref{meff}) in Appendix~\ref{appA}, reads $m^*=0.38m_0$ with $m_0$ being the
free electron mass. The coefficients of the SOC $\lambda_c=1.5\ $meV,\cite{kadantsev152,kosmider87,kormanyos} 
$A_1=417.94\ $meV\ \r{A}$^2$ and $A_2=92.52\ $meV\ \r{A}$^3$, in which $A_1$ and $A_2$ are calculated according to
  Eqs.~(\ref{A_1})-(\ref{A_2}). The other
parameters related to the two-band Hamiltonian [Eq.~(\ref{totalhamil})] and scattering matrix elements are given in 
Ref.~\onlinecite{rostami} and Ref.~\onlinecite{wangarXiv}, respectively. Then we present our results
by numerically solving the KSBEs [Eq.~(\ref{KSBE})].\cite{wureview} 
The initial spin polarization is taken to be $2.5$~\% and the spin-polarization 
direction is set to be in the monolayer MoS$_2$ plane unless otherwise specified.

\begin{figure}[bth]
\centering
\includegraphics[width=8.5cm]{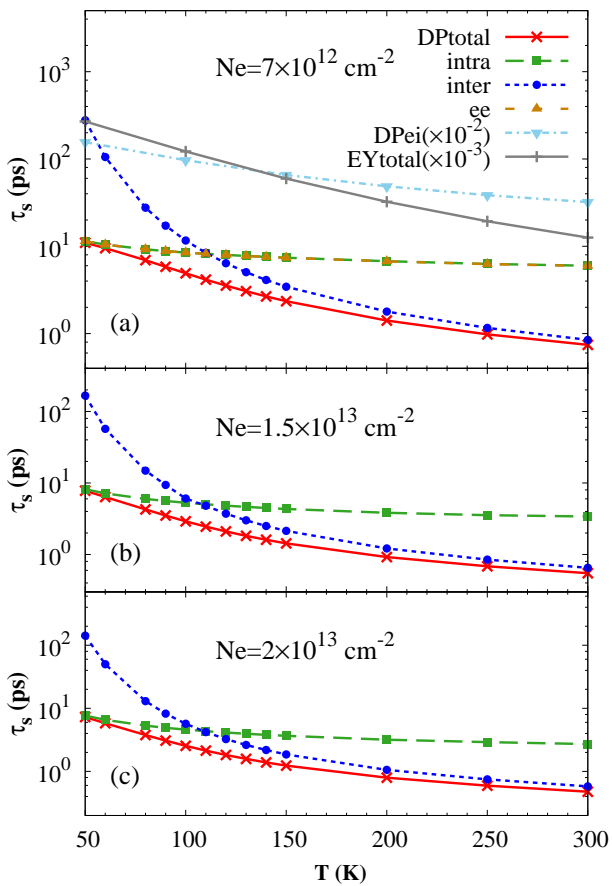}
\caption{(Color online) Total in-plane SRT $\tau_{s}$ ($\times$) due to the DP
  mechanism and that calculated with only the intravalley ($\blacksquare$) or
  intervalley process ($\bullet$) included as function of temperature
  $T$ with (a) $N_e=7\times 10^{12}\ $cm$^{-2}$; (b) $N_e=1.5\times 10^{13}\
  $cm$^{-2}$; and (c) $N_e=2\times 10^{13}\ $cm$^{-2}$. In addition, in (a),
  curve with $\blacktriangle$ ($\blacktriangledown$) stands for $\tau_{s}$ due
to the DP mechanism with only the electron-electron Coulomb (electron-impurity) 
scattering whereas the one with $+$ represents the SRT due
to the EY mechanism. The impurity density $N_i=0.1N_e$.}
\label{fig1}
\end{figure}

We first investigate the temperature dependence of the in-plane spin relaxation.
The in-plane SRTs $\tau_{s}$ as function of temperature $T$ with different
electron densities are plotted in Fig.~\ref{fig1}. 
The impurity density is taken to be $N_i=0.1N_e$ with $N_e$
  being the electron density. We first focus on the case of
$N_e=7\times 10^{12}\ $cm$^{-2}$, i.e., Fig.~\ref{fig1}(a). By
comparing the EY (curve with $+$) and DP
 (curve with $\times$) SRTs, we find that the SRT
due to the EY mechanism is about four orders of magnitude larger than the one
due to the DP mechanism. It is noted that when we artificially increase the
  impurity density to $N_i=10N_e$, the EY SRT is still two orders of magnitude larger
  than the DP one (not shown). This indicates that the contribution of the EY
mechanism to the in-plane spin relaxation is negligible, which
originates from the extremely small spin mixing as shown
in Appendix~\ref{appA}. We also find that the dominant DP SRT presents a monotonic decrease with the
increase of the temperature. To understand this behavior, we calculate the SRT
due to the DP mechanism with only the intravalley ($\blacksquare$) or intervalley
($\bullet$) process included, separately. It is seen that
both the intra- and inter-valley\cite{wangarXiv} SRTs decrease with
increasing temperature, leading to the decrease of the total SRT. 
By comparing these two processes, the intravalley one is found to be more important at low
temperature whereas the intervalley one plays a more important role at high
temperature.

In addition, in contrast to the rapid decrease of the SRT of the intervalley
process,\cite{wangarXiv} we find that the intravalley SRT decreases
mildly with the increase of the temperature when the electron density $N_e=7\times 10^{12}\ $cm$^{-2}$
[see Fig.~\ref{fig1}(a)]. Similar behaviors are also observed when we increase the
electron density to $N_e=1.5\times 10^{13}\ $cm$^{-2}$ and $N_e=2\times
10^{13}\ $cm$^{-2}$ as shown in Figs.~\ref{fig1}(b) and \ref{fig1}(c), respectively. To facilitate
the understanding of this behavior, we calculate the intravalley SRT 
with only the electron-electron Coulomb, electron-impurity, or intravalley
electron-phonon scattering included, separately. We find that the
  intravalley process is dominated by the electron-electron Coulomb
  scattering [only show the
case of $N_e=7\times 10^{12}\ $cm$^{-2}$ in Fig.~\ref{fig1}(a)],
which is very different from the previous studies in
  semiconductors\cite{wureview} and graphene\cite{yzhou,lwang} where
  the electron-impurity scattering plays a very important role in spin
relaxation. Here, the
marginal contribution of the electron-impurity scattering to the
intravalley DP spin relaxation is due to the 
extremely week inhomogeneous broadening from 
the anisotropic cubic term in the SOC [see
Eq.~(\ref{soc})]. If this cubic term is further neglected, the DP spin
relaxation due to the electron-impurity scattering becomes absent as
previously mentioned.

The decrease of the intravalley SRT due to the electron-electron
Coulomb scattering with increasing temperature in the
  degenerate limit\cite{deg} is very different from the previous
  studies in
  semiconductors,\cite{jzhou,leyland,ruan,zhang80,yzhou11,bysun,lfhan,lwang111} where 
a peak in the temperature dependence of the SRT
from the degenerate-to-nondegenerate 
limit was theoretically predicted\cite{jzhou,jiang}
and experimentally realized\cite{leyland,ruan,lfhan} when the spin relaxation is dominated by
the electron-electron Coulomb scattering. 
The underlying physics can be understood from a simplified two-state model
detailed in Appendix~\ref{appB}. When the system is highly degenerate, the
inhomogeneous broadening from the second term of
the SOC in Eq.~(\ref{soc}) is proportional to $T^2$, which suppresses the enhancement of the
electron-electron Coulomb scattering from the weakened Pauli blocking with
increasing temperature.\cite{glazov} This leads to the decrease of the SRT with the
increase of temperature. It is noted that when we
artificially neglect the polar angle dependence of the momentum in the SOC [similar to
the second term of the SOC in Eq.~(\ref{soc})] in semiconductors, the
SRT also decreases with increasing temperature in the degenerate
limit and the peak in the temperature dependence of the SRT becomes
absent too (not shown).

\begin{figure}[bth]
\centering
\includegraphics[width=8.5cm]{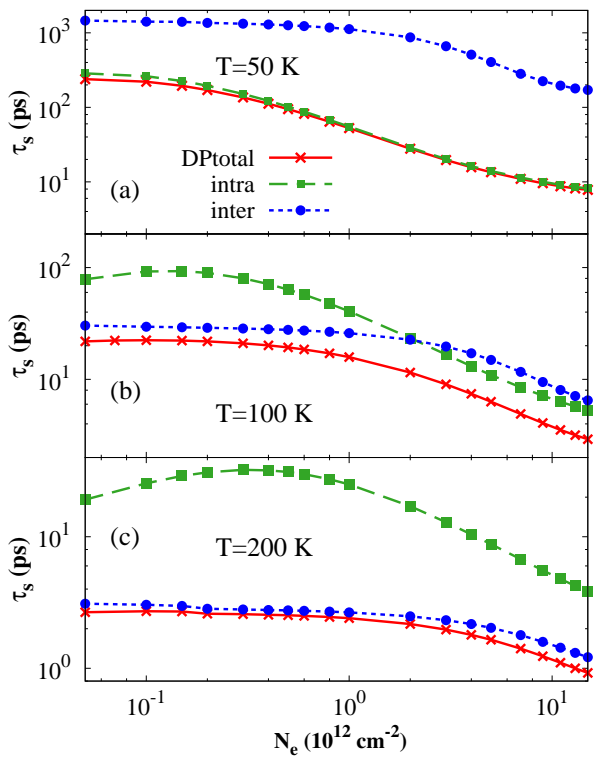}
\caption{(Color online) Total in-plane SRT $\tau_{s}$ ($\times$) due to the DP
  mechanism and that calculated with only the intravalley ($\blacksquare$) or
  intervalley process ($\bullet$) included as function of the
  electron density $N_e$ at (a) $T=50\ $K; (b) $T=100\ $K; and (c) $T=200\ $K. 
The impurity density $N_i=0.1N_e$.}
\label{fig2}
\end{figure}

Then we turn to study the dependence of the in-plane SRT on the electron
density. As the contribution of the EY mechanism to the spin relaxation is
negligible, we only calculate the total SRT and its intravalley or intervalley
process due to the DP mechanism. With the impurity density
$N_i=0.1N_e$, the results at different temperatures are
plotted in Fig.~\ref{fig2}. We
find that at low temperature (i.e., $T=50\ $K), the intravalley
process is always more important than the intervalley one as the
electron density increases. However, 
at high temperature (i.e., $T=200\ $K), the intervalley process
becomes more important. As for the temperature in between (i.e.,
$T=100\ $K), the leading process is
found to change from the intervalley to intravalley one with the
increase of electron density. In addition, we find that the
  total SRTs in different cases all decrease with the
  increase of electron density. We also find that
  the intervalley SRTs show a monotonic decrease with increasing
  electron density. This is because the intervalley electron-phonon scattering is in the
  weak scattering limit, which determines the decrease of the intervalley SRT
  with the enhancement of the intervalley scattering as the electron density
  increases.\cite{wangarXiv} As for the
  intravalley process, the SRT decreases with increasing electron
  density at $T=50\ $K whereas a peak is observed in the density
  dependence of the SRT at $T=100$ and $200\ $K due to the
  crossover from the nondegenerate-to-degenerate
  limit.\cite{jiang,kraub} The increase of the intravalley SRT at
  $T=100$ and $200\ $K is suppressed by
  the decrease of the intervalley one, which leads to the decrease of
  the total SRT with increasing electron density.

\section{Summary}
In summary, we have investigated the in-plane spin relaxation of
electrons in monolayer MoS$_2$. We construct the effective Hamiltonian for
  the conduction band by the L\"{o}wdin partition method 
including the anisotropic two-band Hamiltonian and the intrinsic SOC of the conduction band. We
find that the SOC of the conduction band can lead to the
intra- and inter-valley DP spin relaxation. In addition, the EY spin
relaxation also exists due to the interband spin mixing. 

We calculate the electron spin relaxation due to the DP and EY
mechanisms including the intra- and inter-valley processes by
numerically solving the KSBEs with all the relevant scatterings
included. We find that the in-plane spin relaxation is dominated by the DP
mechanism whereas the contribution of the EY mechanism is marginal due to the
extremely weak in-plane spin mixing. For the dominant DP mechanism, the intravalley process
is found to be more important at low temperature whereas the
intervalley one plays a more important role at high temperature. At the
temperature in between, a crossover of the leading process from the intervalley to intravalley
one is shown with the increase of electron density. Moreover, 
we find that even in the presence of high impurity density, the
intravalley process is dominated by the electron-electron Coulomb
scattering since only the negligible anisotropic term in the SOC
contributes to the intravalley spin relaxation due to the electron-impurity
scattering. This is quite 
different from semiconductors and graphene. 
In addition, the decrease of the intravalley SRT due to the
electron-electron Coulomb scattering with the increase of temperature
in the degenerate limit is of great difference from the previous studies in
semiconductors, where a peak in the temperature dependence of the SRT
was theoretically predicted and experimentally realized when the
electron-electron Coulomb scattering is dominant.

\begin{acknowledgments}
This work was supported by the National Natural Science Foundation of
China under Grant No.\ 11334014, the National Basic Research Program of
China under Grant No.\ 2012CB922002 and the Strategic 
Priority Research Program of the
Chinese Academy of Sciences under Grant No.\ XDB01000000. 
\end{acknowledgments}

\begin{appendix}
  \section{EFFECTIVE HAMILTONIAN AND SPIN MIXING  $\hat{\Lambda}_{\mu{\bf k},{\mu}^{\prime}{\bf
    k}^{\prime}}$ FOR THE CONDUCTION
      BAND AND THE ELECTRON-ELECTRON COULOMB SCATTERING TERM}\label{appA}
Including the anisotropic two-band Hamiltonian [Eq.~(\ref{totalhamil})] and the
intrinsic SOC of the conduction band, the total two-band Hamiltonian
describing the low-energy conduction and valence bands
near the K(K$^{\prime}$) point reads $H^{\mu}_{\rm tot}=H^{\rm a}_{\mu}+\lambda_c\mu \sigma_z(1+\tau_z)/{2}$. 
We define the leading part of this Hamiltonian (i.e., momentum-independent part) as 
$H_0=\Delta\tau_z/2+\lambda_v\mu \sigma_z(1-\tau_z)/{2}+\lambda_c\mu
\sigma_z(1+\tau_z)/{2}$. By considering
the large energy gap $\Delta$, we construct the effective Hamiltonian
of the conduction band by the L\"{o}wdin partition method.\cite{lowdin,winkler} Up to the
third order of the momentum, the effective Hamiltonian is given in
Eqs.~(\ref{soc}) and (\ref{hamil}). The effective mass $m^*$ in
Eq.~(\ref{hamil}) and the coefficients of the SOC $A_1$, $A_2$ in Eq.~(\ref{soc}) read
\begin{eqnarray}
{m^*}^{-1}&=&{2a_0^2t_0^2}/({\hbar^2\Delta})+({\alpha+\beta})/{(2m_0)},\label{meff}\\
A_1&=&{2a_0^2t_0^2(\lambda_v-\lambda_c)}/{\Delta^2},\label{A_1}\\
A_2&=&{4t_0t_1a_0^3(\lambda_v-\lambda_c)}/{\Delta^2}.\label{A_2}
\end{eqnarray}

In addition, the spin mixing $\hat{\Lambda}_{\mu{\bf k},{\mu}^{\prime}{\bf
    k}^{\prime}}$ for the conduction band in
the spin-flip scattering due to the EY mechanism is given by
$\hat{\Lambda}_{\mu{\bf k},{\mu}^{\prime}{\bf
    k}^{\prime}}=\hat{I}-[S_{\mu{\bf k}}^{(1)}{S_{\mu{\bf
      k}}^{(1)}}^{\dagger}-2S_{\mu{\bf k}}^{(1)}{S_{{\mu}^{\prime}{\bf
      k}^{\prime}}^{(1)}}^{\dagger}+S_{{\mu}^{\prime}{\bf k}^{\prime}}^{(1)}{S_{{\mu}^{\prime}{\bf
      k}^{\prime}}^{(1)}}^{\dagger}]/2$ with $\hat{I}$ standing for a $2\times 2$ unit
matrix. $S_{\mu{\bf k}}^{(1)}$ can be written as
\begin{eqnarray}
S_{\mu{\bf k}}^{(1)}&=&-\{a_0t_0(\mu k_x-ik_y)+t_1a_0^2[(k_x^2-k_y^2)+2i\mu
k_xk_y]\}\nonumber\\
&&\mbox{}\times[{\hat{I}}/{\Delta}+{\mu
  \sigma_z(\lambda_v-\lambda_c)}/{\Delta^2}]. 
\end{eqnarray}
It is noted that the unit matrix in $\hat{\Lambda}_{\mu{\bf
    k},{\mu}^{\prime}{\bf k}^{\prime}}$ does not cause any spin flipping. 
The second term $S_{\mu{\bf k}}^{(1)}{S_{\mu{\bf k}}^{(1)}}^{\dagger}$ is
proportional to $[\frac{\hat{I}}{\Delta}+\frac{\mu
  \sigma_z(\lambda_v-\lambda_c)}{\Delta^2}]^2=[\frac{1}{\Delta^2}+\frac{(\lambda_v-\lambda_c)^2}{\Delta^4}]\hat{I}
+\frac{2(\lambda_v-\lambda_c)\mu}{\Delta^3}\sigma_z$ where only the term containing
$\sigma_z$ induces the spin flipping for the in-plane spins. However, this term ($\propto \Delta^{-3}$) is
negligible due to the large energy gap. Similarly, the contribution of the last two
terms is also marginal. 

With the spin mixing $\hat{\Lambda}_{\mu{\bf k},{\mu}^{\prime}{\bf
    k}^{\prime}}$ included, the electron-electron Coulomb scattering term in
Eq.~(\ref{KSBE}) can be written as
\begin{eqnarray}
 \partial_{t}{\rho}_{{\mu}{\bf k}}|_{\rm ee} &=&-\pi\sum_{\mu^\prime{\bf k}^{\prime}{\bf k}^{\prime\prime}}
  |V^{\mu}_{{\bf k},{\bf k}^{\prime}}|^2\delta(\epsilon_{{\mu}{\bf k}^{\prime}}-\epsilon_{{\mu}{\bf k}}
+\epsilon_{{\mu}^\prime{\bf k}^{\prime\prime}}\nonumber\\
&&\hspace{-1cm}\mbox{}-\epsilon_{\mathtt{\mu^\prime{\bf k}^{\prime\prime}-{\bf k}+{\bf k}^{\prime}}})
\Big[{\rm Tr}(\hat{\Lambda}_{\mu^{\prime}{\bf k}^{\prime\prime},{\mu}^{\prime}{\bf k}^{\prime\prime}{-}{\bf k}{+}{\bf k}^{\prime}}{\rho}^{<}_{\mu^\prime{\bf k}^{\prime\prime}{-}{\bf k}{+}{\bf k}^{\prime}}\nonumber\\
&&\hspace{-1cm}\mbox{}\times\hat{\Lambda}_{{\mu}^{\prime}{\bf k}^{\prime\prime}{-}{\bf k}{+}{\bf k}^{\prime},\mu^{\prime}{\bf k}^{\prime\prime}}{\rho}^{>}_{\mu^\prime{\bf k}^{\prime\prime}})\hat{\Lambda}_{\mu{\bf k},{\mu}{\bf
    k}^{\prime}}{\rho}^{>}_{\mu{\bf k}^{\prime}}\hat{\Lambda}_{\mu{\bf k}^{\prime},{\mu}{\bf
    k}}{\rho}^{<}_{\mu{\bf k}}\nonumber\\
&&\hspace{-1cm}\mbox{}-{\rm Tr}(\hat{\Lambda}_{\mu^{\prime}{\bf k}^{\prime\prime},{\mu}^{\prime}{\bf k}^{\prime\prime}{-}{\bf k}{+}{\bf k}^{\prime}}{\rho}^{>}_{\mu^\prime{\bf k}^{\prime\prime}{-}{\bf k}{+}{\bf k}^{\prime}}\hat{\Lambda}_{{\mu}^{\prime}{\bf k}^{\prime\prime}{-}{\bf k}{+}{\bf k}^{\prime},\mu^{\prime}{\bf k}^{\prime\prime}}\nonumber\\
&&\hspace{-1cm}\mbox{}\times{\rho}^{<}_{\mu^\prime{\bf
    k}^{\prime\prime}})\hat{\Lambda}_{\mu{\bf k},{\mu}{\bf
    k}^{\prime}}{\rho}^{<}_{\mu{\bf
    k}^{\prime}}\hat{\Lambda}_{\mu{\bf k}^{\prime},{\mu}{\bf
    k}}{\rho}^{>}_{\mu{\bf k}}\Big]+{\rm H.c.}, 
\end{eqnarray}
with $|V^{\mu}_{{\bf k},{\bf k}^{\prime}}|^2$ being the
electron-electron Coulomb scattering matrix element.

\section{A SIMPLIFIED TWO-STATE MODEL}\label{appB}
We consider the intravalley in-plane spin relaxation due to the DP mechanism with only the electron-electron Coulomb
scattering included. The KSBEs, i.e., Eq.~(\ref{KSBE}) can be written as\cite{negcubic} 
\begin{eqnarray}
\frac{\partial \rho_{\bf k}}{\partial
  t}+\frac{i}{2\hbar}[(2\lambda_c+A_1k^2)\sigma_z,\rho_{\bf k}]=-\sum_{{\bf
    k}^{\prime}}\frac{\rho_{\bf k}-\rho_{{\bf k}^{\prime}}}{\tau^{\rm ee}_{|{\bf k}-{\bf k}^{\prime}|}},
\end{eqnarray}
with the relaxation time approximation where $\tau^{\rm ee}_{|{\bf k}-{\bf k}^{\prime}|}$ represents the momentum scattering
time due to electron-electron Coulomb scattering. It is noted that the valley
index $\mu$ is omitted since only the intravalley process is considered here.
As it is difficult to solve this equation analytically, we
start from a two-state model for simplicity. Specifically, 
\begin{eqnarray}
\frac{\partial \rho_{k}}{\partial
  t}+\frac{i}{2\hbar}[(2\lambda_c+A_1k^2)\sigma_z,\rho_{k}]&=&-\frac{\rho_{k}-\rho_{{k}^{\prime}}}{\tau^{\rm
  ee}},\\
\frac{\partial \rho_{{k}^{\prime}}}{\partial
  t}+\frac{i}{2\hbar}[(2\lambda_c+A_1{k^{\prime}}^2)\sigma_z,\rho_{{k}^{\prime}}]&=&-\frac{\rho_{{k}^{\prime}}-\rho_{k}}{\tau^{\rm
  ee}},
\end{eqnarray}  
It is noted that the isotropic approximation is employed by considering the isotropy of the SOC.\cite{negcubic}
With $B_1=2\lambda_c+A_1(k^2+{k^{\prime}}^2)/2$, $B_2=A_1(k^2-{k^{\prime}}^2)/2$
and unitary transformation $\rho_k=e^{-iB_1\sigma_z
  t/(2\hbar)}\tilde{\rho}_ke^{iB_1\sigma_z t/(2\hbar)}$, the above equations
become
\begin{eqnarray}
\frac{\partial\tilde{\rho}_{{k}}}{\partial
  t}+\frac{i}{2\hbar}B_2[\sigma_z,\tilde{\rho}_{{k}}]=-\frac{\tilde{\rho}_k-\tilde{\rho}_{k^{\prime}}}{\tau^{\rm
  ee}},\\
\frac{\partial\tilde{\rho}_{{k}^{\prime}}}{\partial
  t}-\frac{i}{2\hbar}B_2[\sigma_z,\tilde{\rho}_{{k}^{\prime}}]=-\frac{\tilde{\rho}_{k^{\prime}}-\tilde{\rho}_k}{\tau^{\rm
  ee}}. 
\end{eqnarray}   
By defining the spin vector as $\tilde{\bf S}_k(t)={\rm
  Tr}(\tilde{\rho}_{{k}}{\bgreek \sigma})$, one obtains 
\begin{eqnarray}
&&\frac{\partial \tilde{\bf S}_{k}}{\partial
  t}+\frac{B_2}{\hbar}(\tilde{\bf S}_{k}\times \hat{\bf z})=-\frac{\tilde{\bf
    S}_{k}-\tilde{\bf S}_{k^{\prime}}}{\tau^{\rm
  ee}},\\
&&\frac{\partial \tilde{\bf S}_{k^{\prime}}}{\partial
  t}-\frac{B_2}{\hbar}(\tilde{\bf S}_{k^{\prime}}\times \hat{\bf z})=-\frac{\tilde{\bf
    S}_{k^{\prime}}-\tilde{\bf S}_{k}}{\tau^{\rm
  ee}}.
\end{eqnarray}  
With the initial spin polarization along $x$-axis $\tilde{S}^x_{k}(0)=P_{k0}$,
$\tilde{S}_{k^{\prime}}^x(0)=P_{k^{\prime}0}$, and $\tilde{S}_{k}^y(0)=\tilde{S}_{k^{\prime}}^y(0)=0$, we have 
\begin{eqnarray}
\tilde{S}_{k}^x+\tilde{S}_{k^{\prime}}^x&=&(P_{k0}+P_{k^{\prime}0})e^{-B_2^2\tau^{\rm
  ee}
  t/(2\hbar^2)},\\
\tilde{S}_{k}^y+\tilde{S}_{k^{\prime}}^y&=&(P_{k0}-P_{k^{\prime}0})[e^{-B_2^2\tau^{\rm ee}t/(2\hbar^2)}\nonumber\\
&&\mbox{}-e^{-2t/\tau^{\rm ee}}]B_2\tau^{\rm ee}/(2\hbar),
\end{eqnarray}    
by considering $|B_2|\tau^{\rm ee}/\hbar\ll 1$. Then, the total spin
vector along the $x$-direction reads 
$S_{k}^x+S_{k^{\prime}}^x={\rm
  Tr}[(\rho_k+\rho_{k^{\prime}})\sigma_x]=\cos(B_1t/\hbar)(\tilde{S}_{k}^x
+\tilde{S}_{k^{\prime}}^x)-
\sin(B_1t/\hbar)(\tilde{S}_{k}^y+\tilde{S}_{k^{\prime}}^y)\approx
(P_{k0}+P_{k^{\prime}0})\cos(B_1t/\hbar)e^{-tB_2^2\tau^{\rm ee}/(2\hbar^2)}$. In
the degenerate limit, one obtains
\begin{eqnarray}
S_{k_F}^x\approx P_{k_F0}\cos[(2\lambda_c+A_1k_F^2)t/\hbar]e^{-t/\tau_s(k_F)},
\end{eqnarray}
with the SRT $\tau_s(k_F)={2\hbar^6}/({A_1^2{m^*}^2k^2_BT^2\tau^{\rm
    ee}_{k_F}})$ by considering $|\epsilon_{k}-\epsilon_{k^{\prime}}|\sim
k_BT$. Here, $k_F$ and $k_B$ represent the Fermi wave vector and the Boltzmann constant, respectively. 
It is noted that the inhomogeneous broadening [i.e.,
${A_1^2{m^*}^2k^2_BT^2}/({2\hbar^6})$] is proportional to $T^2$. In
addition, $1/\tau_{k_F}^{\rm ee}\propto \ln(E_F/k_BT)T^2/E_F$ in the degenerate
limit with $E_F$ being the Fermi energy.\cite{glazov} Therefore, one has $\tau_s(k_F)\propto \ln(E_F/k_BT)/E_F$. 
\end{appendix}


\begin{thebibliography}{0}
\bibitem{warping} It is noted that the trigonal warping term (i.e., spinless cubic term) is neglected since
  its contribution to the energy of the conduction band is marginal.
\bibitem{scat} Note that the scatterings in Ref.~\onlinecite{jiang} are in
bulk system and one has to transform them to
two-dimensional case. In addition, one also has to transform these scatterings
from single valley to two valleys, which can refer to Ref.~\onlinecite{yzhou}.
\bibitem{negcubic} It is noted that we neglect the cubic term of the SOC since
  its contribution to the intravalley spin relaxation is marginal when only the
  electron-electron Coulomb scattering is included.
\end{thebibliography}
\end{document}